\documentclass[preprint,12pt,authoryear]{elsarticle}

\usepackage{amssymb}
\usepackage{url}
\usepackage{booktabs} 
\usepackage{csquotes}
\usepackage{enumitem}  
\usepackage{xcolor} 
\usepackage{multirow} 
\usepackage{dirtytalk}
\usepackage{caption}
\usepackage{xcolor}
\newcommand{\lsq}[1]{\textcolor{black}{#1}} 
 
\usepackage{tabularx}
\journal{HBET}

\begin{document}

\begin{frontmatter}




\title{The Impact of AI Generated Content on Decision Making for Topics Requiring Expertise}


\author[inst1]{Shangqian Li\footnote{Email address: uqsli24@uq.edu.au (Shangqian Li)}}
\author[inst1]{Tianwa Chen}
\author[inst1]{Gianluca Demartini}
\affiliation[inst1]{organization={School of Electrical Engineering and Computer Science, University Of Queensland},
            addressline={St Lucia}, 
            city={Brisbane},
            state={Queensland},
            country={Australia}}

\newpageafter{author}




\begin{abstract}
Modelling users' online decision-making and opinion change is a complex issue that needs to consider users' personal determinants, the nature of the topic and the information retrieval activities.
Furthermore, generative-AI-based products like ChatGPT gradually become an essential element for the retrieval of online information. However, the interaction between domain-specific knowledge and AI-generated content during online decision-making is unclear. We conducted a lab-based explanatory sequential 
study with university students
to overcome this research gap. In the experiment, we surveyed participants about a set of general domain topics that are easy to grasp and another set of domain-specific topics that require adequate levels of chemical science knowledge to fully comprehend. We provided participants with decision-supporting information that was either produced using generative AI or collected from selected expert human-written sources to explore the role of AI-generated content compared to ordinary information during decision-making. Our result revealed that participants are less likely to change opinions on domain-specific topics. Since participants without professional knowledge had difficulty performing in-depth and independent reasoning based on the information, they favoured relying on conclusions presented in the provided materials and tended to stick to their initial opinion. Besides, information that is labelled as AI-generated is equivalently helpful as information labelled as dedicatedly human-written for participants in this experiment, indicating the vast potential as well as concerns for AI replacing human experts to help users tackle professional topics or issues. 
\end{abstract}




\end{frontmatter}


\section{Introduction}

Opinion dynamics refers to individuals changing their thoughts or attitudes regarding an 
issue after receiving new information or signals from others. It can take place anytime online, as long as users keep viewing new content. Since Computer-Mediated Channels (CMCs), such as social media and online forums, replaced traditional media channels like radio and television,  public's daily information retrieval habits were reshaped \citep{RN302}. CMCs allow for more frequent consumption of new information \citep{RN303}. Moreover, CMCs allow users to quickly exchange messages and affect each other's opinion formation processes \citep{RN303}. This feature brings up the benefit of encouraging users to access diversified ideas and viewpoints. However, the drawback is also significant. For example, the increased difficulty in auditing the veracity of User-Generated Content (UGC) and assessing the reliability of AI-Generated Content (AIGC) would inevitably cause mis- and disinformation spread issues \citep{RN113}.

Existing research has observed that understanding  individual's decision-making process is essential to tackle the misinformation spread issue \citep{RN304}. Nevertheless, a range of studies on online opinion dynamics and conformity behaviours have reached a consent that modelling users' thought is challenging because (i) opinions are difficult to capture unless samples explicitly state their considerations \citep{RN115}, (ii) subjects could have issues expressing feelings due to subjective bias or external factors such as social pressure and social norms \citep{RN2, RN103}, (iii) it is challenging to measure the effects of numerous confounding factors in real-life Social Networking Sites (SNS) environment, for instance – the impact of social circles \citep{RN103, RN49}.

Observations advise that users' opinions also fundamentally depend on the signal received, including the perceived trustworthiness of information sources (e.g., arbitrary SNS users and fact-checking sites) and other users' participation (e.g., upvotes, likes and shares) \citep{RN88,  RN91}. So far, we have insufficient understanding about how AI-related signals would affect one's information preference concerning decision-making. Furthermore, it was stated that personal determinants contribute to users' preference for information that may affect their final decision \citep{RN103, RN49}. However, concluding the correlation between topic-related domain-specific knowledge and information preference in statements is challenging because of the lack of work quantifying the explicit connection among domain-specific knowledge levels, information preference and the consequent decision-making. The above gaps motivated us to conduct this research.
Thus, in this work we address the following research questions (RQs):

\textit{\textbf{RQ1: How does users' domain-specific knowledge and self-recognised expertise affect their decision-making?}}

\textit{\textbf{RQ2: How does users' perceptions of AIGC affect their decision-making?}}

We adopted an explanatory sequential research design \citep{RN300} to answer the above research questions. 
Accordingly, a survey-based experiment was developed, where participants were asked to complete eight tasks, each following the same procedure. 
The survey concluded once all tasks were submitted. Immediately following the survey, we conducted one-on-one semi-structured interviews with each participant to further explore Research Questions 1 and 2.

\section{Related Work}
\subsection{Perceived information-related signals}

How the  informativeness of the message affect subjects' decision-making primarily depends on the topic \citep{RN36}. According to \citet{RN305} and \citet{RN332}'s investigation, individuals' knowledge regarding the topic of discussion is determinative. Individuals who know little about the task would have difficulty critically evaluating messages they receive and consequently would rely more on the perceived signals to assist judgements \citep{RN331}. In contrast, if no obstacle hinders individuals from elaborating on the topic, the informativeness of the messages retained would weigh more during the decision-making \citep{RN103}. Unfortunately, the above findings make the effect of informational influence controversial. During the debate of political elections, most online participants have explicit opinions of ongoing matters \citep{RN6}. Therefore, theoretically, individuals would place the informativeness-related factors in advance due to their high domain-specific knowledge. However, evidence indicates that information signals are still one of the primary influencers affecting people's decision-making in real-life SNS, even if participants associated with opinion deliberations are highly familiar with and relevant to the topic \citep{RN6, RN14}. The primary reason for this conflict could be the differences in the tasks. In most experiment-based studies, participants are given a specified genre of tasks or filtered topics for discussions \citep{RN2, RN3, RN49, RN103}. In contrast, in real-life SNS, participants would encounter a more diverse scope of social events or news that requires different knowledge or causes potential conflicts of interest.

Many of the existing determinations regarding information signals are instinctive. For instance, individuals are more likely to conform to sources that are highly reputed or authoritative \citep{RN307}, and the advancing number of indicators such as comments, shares and likes could create significant senses of the majority group and thus be more persuasive to subjects \citep{RN306}. \citet{RN124} believe the participant's demographic factors, especially their educational level and domain-specific knowledge, are responsible for their corresponding information preference and retrieval behaviours. Experts with remarkable understandings of the topic are considered more judgemental on the decision-supporting messages, which results in a lower likelihood of conformity behaviours and less reliance on the information-related signals. However, this conclusion is drawn from studies with limited sample sizes \citep{RN74}. There is still a high chance that the conclusion does not hold in real-life SNS platforms during opinion deliberations regarding a large-scale social event. Our work aims to clarify what contributes to people's different information preferences and decision-making criteria. 

\subsection{AI-generated content and decision-making}

The delivery of generative AI products, such as ChatGPT by OpenAI, initiates a new era of information retrieval. Evidence revealed that generative AI can precisely summarise and present text \citep{RN125}. Past work estimated that one of the primary usages of generative AI would be answering knowledge-related queries, which raises concerns of dis- and misinformation spread associated with generative AI \citep{RN127, RN333}. Recent research by \citet{RN128} proved that humans are poor at identifying dis- and misinformation in UGC and AIGC, and humans also have issues distinguishing whether AI generates the text. This conclusion is supported by \citet{RN130}, demonstrating that humans perceive little differences between AIGC and UGC without hints or signals. Therefore, there is no doubt that generative AI would facilitate dis- and misinformation spread in real-life SNS if regulation on AIGC is effective. 

The predecessor of AIGC in the opinion dynamics area is \textit{social bots}. Social bots in SNS are known as agents without the capability to communicate with humans but can automatically distribute human-generated text \citep{RN116}. According to \citet{RN14}, many social bots spreading similar opinions would establish high social presence levels for arbitrary users and demonstrate a sense of mainstream voices because individuals may consider bots other users. However, once individuals are informed about social bots, they are less likely to follow or access the bot's opinions because (i) bots establish undersized social presence levels than real human users, and (ii) many individuals instinctively hold aversion attitudes regarding bots\citep{RN133}. 
However, the existing findings regarding bots and opinion dynamics are unlikely to hold for generative AI systems. First, AIGC presents high-quality text compared to social bots' duplicated and homogeneous contexts. Studies imply that generative AI can offer convincing descriptions or explanations across various areas, usually outperforming non-expert humans who provide the source text for social bots~\citep{RN134}. Secondarily, no substantial evidence indicates that most SNS users hold aversion attitudes against generative AI. In contrast, existing literature shows that the public relies on generative AI products because of their heightened trustworthiness~\citep{RN113}. Thirdly, \citet{RN135} stated that the AIGC is more convincing to insufficiently knowledgeable individuals. The real-life SNS would amplify this phenomenon since users encounter a more comprehensive range of topics and news. Nevertheless, \citet{RN136} suggested that domain-specific experts can tell whether AIGC distributes misinformation if the topic is related to their disciplines. Hence, we aim to investigate how AIGC would influence domain experts' decision-making on domain-specific topics in comparison to non-domain experts.

\section{Study Design}


Based on the research objectives, we purposefully designed the study to
understand the impact of domain-specific knowledge and AIGC on decision-making. 
The data collection for our study is twofold. It includes quantitative analysis based on survey data and qualitative insights obtained through the follow-up semi-structured interviews. This approach enables a deeper interpretation of the quantitative results, thereby enriching our understanding of the data.
The survey was developed using Qualtrics\footnote{For more information, please view https://www.qualtrics.com/}, consisting of a set of Multiple-Choice Questions (MCQs) and Binary-Choice Questions (BCQs). All participants completed the survey and interview in the same lab setting, ensuring consistency in the data collection environment.

\subsection{Participant Recruitment and Ethical Concerns}

All participants in the study were university students. There were no restrictions based on demographic factors like gender and age, but we required specific academic backgrounds to analyze the correlation between domain-specific knowledge and information preference. Specifically, half of the participants were from chemical or environmental engineering disciplines. We did not specify any particular educational requirements for participants, and their academic levels ranged from Bachelor’s degree to PhD students.  Participants were recruited on a voluntary basis, and each was offered \$30 for their time upon completion of the study. The study design was reviewed and approved by the Ethics Committee of authors' institution.  \lsq{The overall experiment duration ranged from 30 to 60 minutes. In total, 30 participants contributed to the study. According to \citet{RN299}, a Generalised Linear Mixed-Effect Model (GLMM) allowed us to assume random effects among samples. It allows us to have 240 samples for statistical modelling, which is a sample size that is sufficient for constructing a GLMM.}


\subsection{Experimental Design}

\subsubsection{Capturing opinion dynamics}\label{subsection_321}

To address the RQs, we measure (i) self-ranked confidence that reflects the level of change of mind and (ii) switch of the viewpoint (from agree to disagree and vice versa) as the two decisive signs for opinion modelling. The above experiment setup simplifies human opinions and enables more statistical analysis methods, such as regressions \citep{RN138}.
We only allow one participant to take the experiment simultaneously. Therefore, we assume minor normative influence and expect informational influence to dominate the subject's conformity behaviour. Furthermore, according to \citet{RN49}, subject bias may dominate the subject's decision-making performance during the task if we inform subjects of the actual experiment purpose before the survey. 


\subsubsection{Topics and domain-specific knowledge}\label{subsection_322}

\citet{RN139} noted that 
deploying socially divisive topics can further trigger the participant's critical thinking and simulate real-life reasoning scenarios. Therefore, in this experiment, we deployed social events from certified news media source, BBC (British Broadcasting Corporation) and ABC (Australia Broadcasting Corporation), to mimic a realistic SNS scenario. Thus, eight topics were drawn from various areas, including politics, technology, science and environment, as shown in Table \ref{tab:topics}. We let participants obtain a regarding the issue by providing them with a short description of about 150 – 300 words. Then we confirm the participant's initial opinion by asking them to answer a BCQ. For instance, if the topic is about banning AIGC for specific purposes, the BCQ would be like, `Do you support banning AIGC for purposes mentioned in the text?'.

\begin{table}
  
  \begin{tabular}{ccl}
    \toprule
      & Sensitive pattern & Insensitive pattern\\
    \midrule
    In- & Bio-disposable plastic & Per- and poly-fluoroalkyl substances (PFAS) \\
    Domain & Carcinogenic chemicals & Ohio chemical leaks\\
    \midrule
    Out-of-  & LSD legalisation & AI-generated art \\
    Domain& Banning TikTok & Drug injection sites \\
    \bottomrule
  \end{tabular}
  \caption{Topic summary}
  \label{tab:topics}
\end{table}

\lsq{In this experiment, we define domain-specific topics as issues associated with chemical engineering (e.g., environmental impacts of chemicals used during fracking and ecological and health impacts of permanent chemicals). As proven by \citet{RN337}, the decision-making procedure on chemical-related topics and problems differs significantly between people with sufficient chemistry background knowledge and people who know little about chemistry. Thus, a chemical background versus a non-chemical background satisfied our assumption that participant benefits from domain-specific knowledge when dealing with domain-specific topics.} We recruited  students from corresponding disciplines, and we thus assume those in-domain students have higher levels of domain-specific knowledge concerning in-domain topics. Both in-domain and out-of-domain participants went through the same set of tasks but in random orders.

\subsubsection{Decision-supporting information and information sources}

For each topic, we prepared five pieces of decision-supporting information for participants to access. Content-wise, each piece of the message can be AIGC or UGC collected from media or relevant websites. We underwent a prompt engineering process to ensure the AIGC obtained shares a similar writing style with the UGC we collected so that participant can hardly distinguish \citep{RN139}. We followed \citet{RN299}'s publication and summarised a list of prompts exploited for generating AIGC, as presented in Table \ref{tab:prompt}. We fact-checked all UGC and AIGC that exploited in this experiment to ensure no mis- or disinformation appears. \lsq{We further ensured the quality of the decision-supporting information by introducing a systematic text-quality evaluation pipeline. A group of domain expert auditors was recruited to assess the quality of the texts. The evaluation group comprised four PhD students—two with backgrounds in chemical engineering and biomedicine—and one postdoctoral researcher. The evaluation metrics covered two primary dimensions: information sufficiency and scientific accuracy. Auditors were required to assign a score ranging from 0 to 5 (where a score above 3.5 indicates a high level of satisfaction) to determine whether (i) the text provided valid information that could assist decision-making, and (ii) the scientific information and knowledge presented were accurate and appropriately conveyed. If auditors were uncertain about judging the scientific accuracy of specific content, they were permitted to skip that portion. Following this systematic evaluation, only decision-supporting texts with an average score above 4 were included in the experiment.}

\begin{table}

  \begin{tabular}{cl}
    \toprule
      & Prompt text\\
    \midrule
    I & Consider the following topic: \\
    &\textit{[input text: a paragraph of the topic or a summary]},\\
    &explain why \textit{[input text: an argument discussed in the }\\
    &\textit{topic introduction]} is true/not true in a paragraph.\\
    & Provide evidence to support your discussion. \\
    \midrule
    II & Read the following argument \textit{[input text: an argument}\\
    &\textit{or discussion]}, and use proper evidence to write a\\ 
    &paragraph to refute/support that argument.\\
    \bottomrule
  \end{tabular}
  \caption{The prompts deployed for generating AIGC}
  \label{tab:prompt}
\end{table}

\lsq{Despite the randomization of topic order, the decision-supporting information within each topic was presented in a fixed sequence. This design choice aimed to minimize potential ambiguity and noise that could be difficult to model (for example, when one participant is exposed to a negative opinion first, whereas another encounters a positive opinion first). To further examine the influence of information order, participants were later invited to discuss in the semi-structured interview how initial opinions shape subsequent decision-making and how the sequence of receiving information with different attitudes affects their judgments.}

Information-related signals, such as the author and format, profoundly affect human cognition \citep{RN91}. In this research, we choose information sources as the primary signal to determine how participants' perceptions of AIGC and information sources counterbalance the information in their decision-making process. After showing them five pieces of decision-supporting information, we ask whether they changed their mind. We then offer them again all the information with references. Subjects are then at the final checkpoint of mind-changing and should note if they alternated opinions at this stage. 

Furthermore, we articulated two patterns to present information sources to determine whether perceptions would result in bias. We only offer the participants with the actual information sources for half of the tasks (source-insensitive pattern), and information sources in the other half will be misleading (source-sensitive pattern), as demonstrated in Table \ref{tab:topics}. In the sensitive pattern, we proactively switched sources of UGC and AIGC, where the sources of all AIGC are presented as UGC and vice-versa.

\subsubsection{Behavioural factors}

In this study, we measured two behavioural factors - (i) thinking time and (ii) number of clicks. Firstly, we exploited JavaScript functions to measure the time subjects spent on each stage and number of clicks on each page. This includes (i) time taken to decide initial, second and final opinion/stance and (ii) time taken to read through each piece of the information \citep{RN118}. When participants access the decision-supporting information, they are allowed to go back and forward to review each information. Thus, the final thinking time and clicks recorded are the total amounts. 

\subsubsection{Participant's attitude towards generative AI}
As  \citet{RN142} concluded, automation and AI are essential information-related signals that could affect people's perceptions of issues and consequently influence their judgments. Therefore, we additionally deployed two methodologies to measure participants' attitudes regarding generative AI - the Trust In Automation questionnaire (TIA) \citep{RN143} and the Affinity for Technology Interaction scale questionnaire (ATI) \citep{RN144}. As introduced by \citet{RN144}, ATI aims to let participants self-rank their perceived enthusiasm for generative AI, the ability to solve technology-related issues and independent learning about technology. In contrast, TIA tends to explore participants' knowledge levels regarding the technology, for instance, the design intention of AI products, participants' proneness to trust such technology, and the technology's reliability and predictability \citep{RN143}. We deploy ATI to estimate participants' subjective perception of AI and exploit TIA to measure their understanding and conviction about AI. The combination of ATI and TIA allows us to gain a thorough overview of the relationship between participants and generative AI.

\subsection{Procedures}

The survey started once we received the consent for the survey and interview recording. We declared that the research question was `How well can generative AI explain emerging issues from the internet'. In this experiment, we asked participants to complete eight identically structured tasks. Tasks are presented in random orders.

At the start of each task, participants read a topic description, followed by completing a set of MCQs to answer: (i) their attitude/stance regarding the topic, (ii) their confidence in their initial decision, (iii) their familiarity with the topic, (iv) the relevance of the topic to themselves, and (v) whether they had previously encountered the topic. The responses to questions (ii), (iii), and (iv) were captured using a Likert scale ranging from 1 to 5, where 1 indicates the lowest level of agreement or familiarity and 5 the highest.
Following this initial assessment, participants were presented with five pieces of decision-supporting information sequentially.

Participants then accessed all five messages. We asked participants (i) any changes in their attitude towards the topic and (ii) their confidence in their decision at that point. We then disclosed the sources of the information and inquired if this revelation led to any change in their views.
Upon completion all tasks, we concluded the survey by revealing (i) the actual research objective and (ii) topics corresponding to source-sensitive and source-insensitive patterns.

The semi-structured interview session commenced once participants were briefed on the experimental setups. The interviews were designed to complement the quantitative findings and delve deeper into  RQ1 and RQ2.
We designed a set of open-ended questions for participants to reflect on their survey responses, including but not limited to: 1.	How does your educational background or knowledge affect the information preference or insights on information sources?
2.	How does the revealing of AIGC and the corresponding prompts influence your decisions?
3. Does prior knowledge of the topic affect your engagement and understanding during information consumption?
4.	Can you explain why you spent more time on particular topics or information?
The interview process is voice recorded and then transcribed.

 \paragraph{Pilot}

Before the main experiment, we conducted a pilot study with six participants (N=6) to ensure the survey content aligned with our experimental assumptions. Based on the feedback received, we revised the decision-supporting information to enhance clarity and ensure participants could easily understand the contents and topics.

\section{Results}

\subsection{Quantitative Results}

\subsubsection{Factors associated with opinion changes}

\begin{figure*}[t]
\centering
\includegraphics[width=1\textwidth]{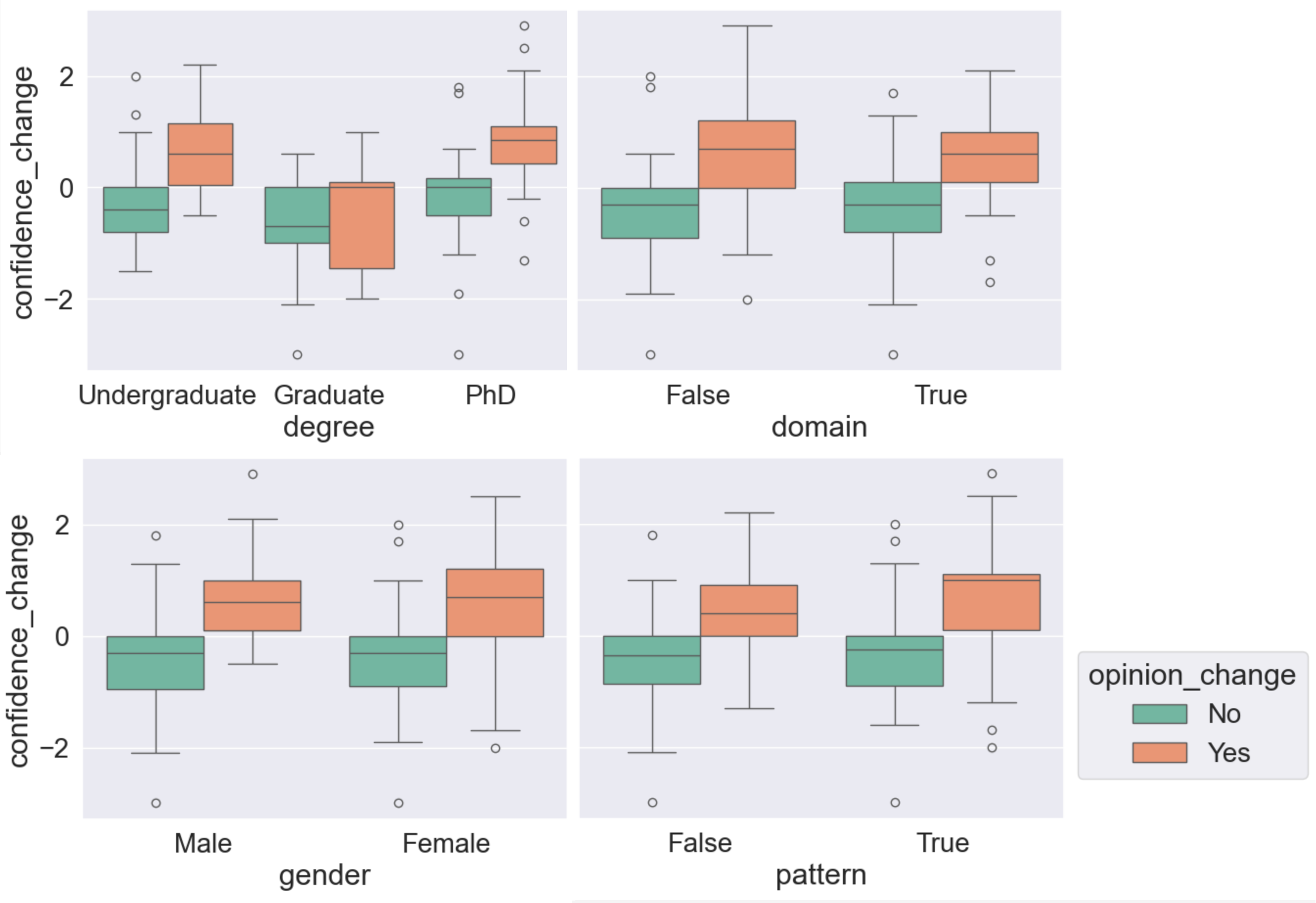} 
\caption{ Plots of confidence changes versus opinion changes under different categorical variables. From left to right: (i) educational degree, (ii) whether the topic is in-domain or out-of-domain, (iii) gender, and (iv) whether the topic is under the source-sensitive pattern or not.}
\label{fig1}
\end{figure*}

\begin{figure*}[!ht]
\centering
\includegraphics[width=1\textwidth]{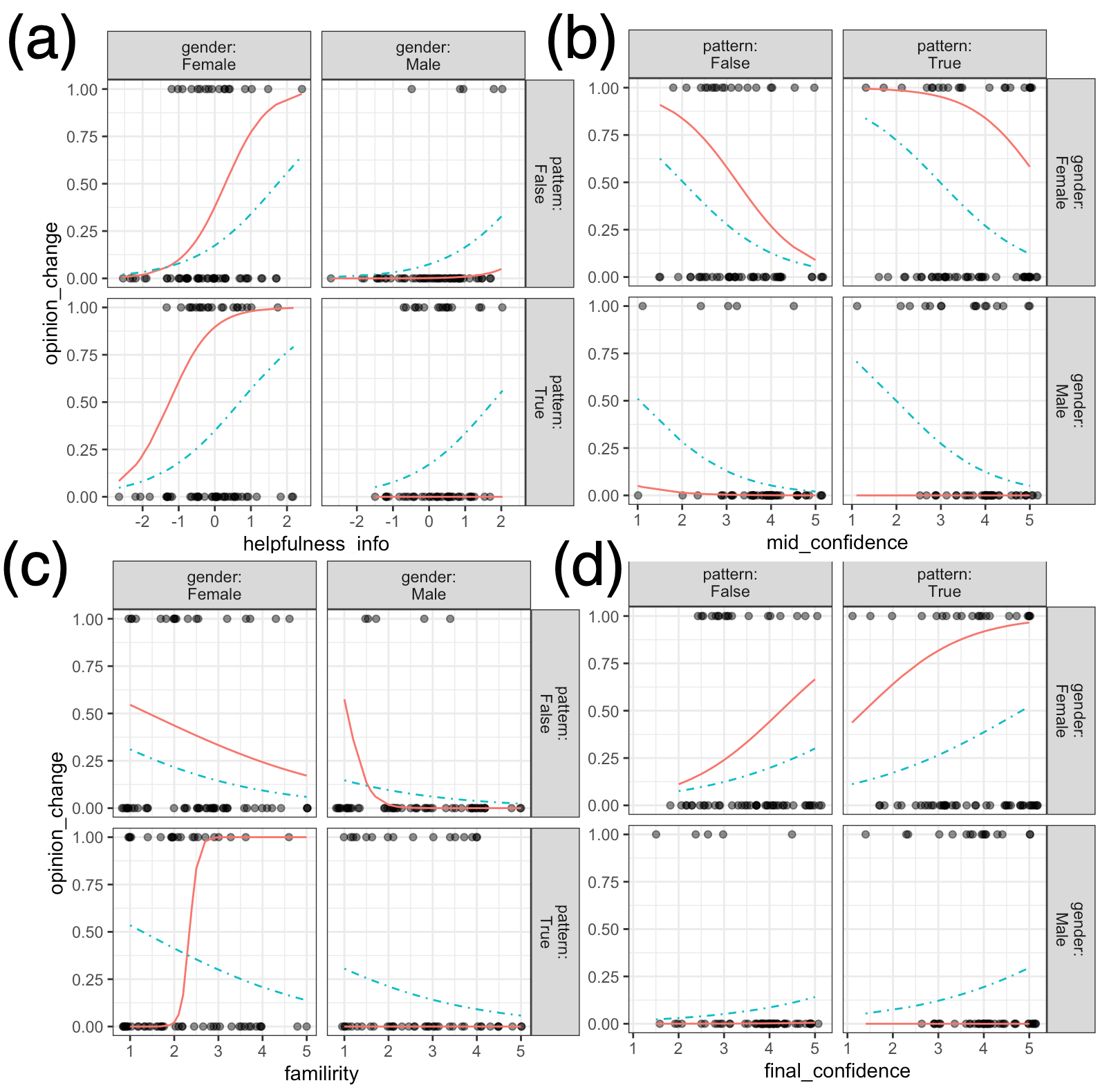} 
\caption{\lsq{Inspection visualisation of the comparison between the full model (red) and the refined model (blue) on opinion changes. It illustrates the combination of gender plus pattern versus different continuous variables. From left to right then top to bottom: (a) perceived helpfulness of the decision-supporting information, (b) confidence before revealing information sources, (c) perceived familiarity regarding the topic, and (d) the final confidence.}}
\label{model1}
\end{figure*}


\begin{table}[t]
  \centering
  \begin{tabular}{ccl}
    \toprule
     \textbf{Dependent Variable} & \multicolumn{2}{c}{opinion\_change} \\
     \midrule
     \textbf{Independent Variable}&&\\
    \midrule
      Fixed Effect & $\beta$ & $p-value$\\
    \midrule
    degreePhD &  2.298 & 0.001 *** \\
    degreeUndergraduate &  1.864 & 0.008 ** \\
    genderMale &  -1.196 & 0.011 * \\
    familiarity  & -0.489 &  0.013 *\\
    final\_confidence & 0.677 & 0.059 .   \\
    patternTrue & 0.807 & 0.037 *  \\
    helpfulness\_info& 0.847& 0.002 **  \\
    mid\_confidence&-1.131 &  0.003 **   \\
    \bottomrule
  \end{tabular}
  \caption{Model summary of the statistically significant predictors for opinion changes}
  \label{tab:glmm_result}
\end{table}

\begin{table}[t]
  \centering

  \begin{tabular}{ccccl}
    \toprule
     \textbf{Model}   &$AIC$&$BIC$ &$Chisq$&$p-value$ \\
     \midrule
    refined.model&225.35 &270.38&&\\
    baseline.model&255.10 &262.02&51.75&  3.026e-07 ***\\
    full.model&290.39& 633.31&106.96&0.043 *\\
    \bottomrule
  \end{tabular}
  \caption{Chi-square-based ANOVA test of the models}
  \label{tab:eval1}
\end{table}

\begin{table}[t]
  \centering

  \begin{tabular}{ccl}
    \toprule
     \textbf{Dependent Variable} &  \multicolumn{2}{c}{confidence\_change} \\ 
     \midrule
     \textbf{Independent Variable}&&\\
    \midrule
      Fixed Effect & $\beta$ & $p-value$\\
    \midrule
    degreePhD &  0.593 & 0.003 **  \\
    degreeUndergraduate &   0.526 & 0.003 **  \\
    familiarity &  0.192 & 3.20e-05 ***  \\
    final\_confidence&-0.695&2.15e-12 ***\\
    final\_time&0.002&0.053 .  \\
    patternTrue&0.219& 0.025 *   \\
    helpfulness\_info&0.156& 0.010 *\\
    mid\_confidence&0.269&0.005 ** \\
    \bottomrule
  \end{tabular}
  \caption{Model summary of the statistically significant predictors for confidence changes}
  \label{tab:glmm_result2}
\end{table}

We model the change in opinion from two perspectives. The first is the self-reported binary change in opinions, and the second is the final self-ranked confidence level (on a [0,1] scale) regarding each decision. In this experiment, 87.27\% of the participants had reported at least one case of changing opinions. 

\lsq{In Fig. \ref{fig1}, we can see that the group of participants who once changed opinions during the experiment were subject to larger confidence changes. PhD students' confidence change shows the largest level of difference and separation between participants who changed their minds and those who did not. The domain plot shows that out-of-domain topics result in a more diverse range of confidence changes regardless of mind changes. We can also observe gender differences since female participants who changed opinions show a wider distribution of confidence change compared to male participants who changed opinions. }

We then exploited \citet{RN299}'s method - a GLMM to determine the statistically significant data attributes contributing to opinion changes and the final confidence level. The results are shown in Table \ref{tab:glmm_result} and \ref{tab:glmm_result2}, where we include the statistically significant factors on the 0.05 and 0.1 levels. To address the assumption of GLMM that each data entry should be independent of the others, we set the respondent ID as the random variable. We deployed a Binomial distribution with a logit link function for the binary change in opinions and a Gaussian distribution with an identity link function for the confidence level. We used \citet{RN202}'s $R^2$ -
$R^2_{GLMM(m)}=\frac{\sigma^2_f}{\sigma^2_f+\sigma^2_a+\sigma^2_{\varepsilon}}$ and $R^2_{GLMM(c)}=\frac{\sigma^2_f+\sigma^2_a}{\sigma^2_f+\sigma^2_a+\sigma^2_{\varepsilon}}$
as an additional layer of goodness-of-fit tests for the models.


Since using a single metric to determine the quality of a binomial GLMM is challenging, we conducted a model selection process to identify the statistically significant factors and obtain the maximum performance model. We proposed a baseline model that only uses intercepts and random effects to predict opinion changes and a full model that uses all available predictors. We start modifying the refined model from the full model. The model selection used the Akaike Information Criterion (AIC) and Bayesian Information Criterion (BIC) as the goodness-of-fit test statistic. After selection, we ran a Chi-square-test-based Analysis of Variance (ANOVA) to determine  significant differences between the three models. The refined model shows the best performance and is distinct from the baseline and the full models, as shown in Table \ref{tab:eval1}.

\lsq{According to the model on opinion change, we identified ten statistically significant predictors. Participants are less likely to change opinions under source-sensitive patterns ($\beta=-0.807$) where we presented wrong information sources. Self-ranked familiarity with the topic is negatively correlated with the likelihood of changing opinion ($\beta=-0.489$). In contrast to self-ranked familiarity, participants' perceived helpfulness of the decision-supporting information ($\beta=0.847$) is positively correlated with the likelihood of changing opinions. That is, the higher the numeric value of the four factors, the higher the odds for a user to change opinions. Besides, participants' confidence levels before and after revealing information sources showed opposite statistical significance. It implied that participants' confidence is negatively correlated with the odds of changing opinions before seeing the information sources ($\beta=-1.131$) but positively correlated with the odds of changing opinions when given information sources ($\beta=0.677$). Furthermore, our model also captured statistical significance regarding participants' degrees and self-reported gender. This indicates that PhD participants ($\beta=2.298$) and undergraduate participants ($\beta=1.864$) are more likely to change their opinions than participants who are currently pursuing a graduate degree. Male participants show a lower likelihood of changing opinions ($\beta=-1.196$). The model has a theoretical $R^2$ of 0.47 and a delta $R^2$ of 0.27, indicating an acceptable level of statistical power. }

\lsq{Furthermore, we obtained eight statistically significant predictors for the confidence level. Similar to the model for binary opinion change, the final confidence and the confidence before obtaining the information source have opposite effects, and graduate participants tend to show lower fluctuation in confidence. Longer thinking time is associated with a larger magnitude of confidence change ($\beta=0.002$). Surprisingly, the information helpfulness ($\beta=0.156$) is also positively correlated with confidence changes, indicating helpful information and high topic relevance could increase the chance of having a larger confidence change. The $R^2_m$ for this model is 0.29, and $R^2_c$ is 0.44, indicating satisfactory statistical power.}

\subsubsection{Breaking down the contextual factors}

We visualised the two models to gain deeper insights into the statistically significant factors. We considered different combinations of continuous variables and categorical variables to examine the interactions. To emphasise the comparison, we used red lines to present the full model and dashed lines to present the refined model for opinion change in Fig. \ref{model1}. 

\lsq{Fig. \ref{model1} takes the two statistically significant categorical variables - gender and pattern. It also determines how each gender and pattern combination interacts with opinion change and other statistically significant predictors. Fig. \ref{model1} (a) shows an evident positive correlation between the likelihood of opinion change and the perceived helpfulness of the decision-supporting information. Female participants show more sensitivity regarding information helpfulness, as the likelihood of changing opinion is higher than that of male participants when the helpfulness is ranked high. In Fig. \ref{model1} (c), the full model and refined model show opposing trends when female participants encountered topics under the source-sensitive pattern. Other than that, the negative correlation between self-rated familiarity regarding the topic and the likelihood of opinion-changing is apparent. Fig. \ref{model1} (b) and (d) is a comparison of the confidence level before and after revealing the information source. Regardless of the gender effect, samples under the source-sensitive pattern show slightly higher sensitivity to opinion changes than the other pattern, according to steeper slopes in Fig. \ref{model1} (b). What's more, similar to Fig. \ref{model1} (a), female participants under the source-sensitivity demonstrate the steepest slope, indicating samples of this combination show the highest likelihood to change opinions when the final confidence is rated high. }

\subsection{Qualitative Results}
We collected qualitative data from two primary sources - the comments participants left during the survey and the semi-structured interview after the survey. 
To analyse the interview data, we follow the approach described by \citep{t:06}. We build themes from the data using thematic analysis (bottom-up approach) to summarise data into abstract codes, themes, or categories. We used an a priori high-level set of codes that aligned with the
four sections of the interview and the research aims. A dual-coder approach was used to reduce any bias in our analysis. Two researchers individually iteratively coded the  22 interviews using the a priori set of codes and creating new related codes as necessary. We used NVivo 12\footnote{NVivo is a qualitative data analysis tool for text-based/or multimedia information. \url{https://help-nv.qsrinternational.com/12/win/v12.1.115-d3ea61/Content/welcome.htm}} to assist in analysing the data and managing the code structure \citep{miles1994qualitative}. The researchers then compared the coding, accounting for synonyms, discussed, and refined the coding until full agreement was reached. Considering the inductive nature of the coding approach, agreement measures were not feasible in the study.

\subsubsection{How do domain-specific knowledge and education degrees affect decision-making? }

In this study, we define (i) in-domain participants as students who have an educational background in chemical engineering (or equivalent fields) and (ii) in-domain topics as topics that necessitate chemical-related knowledge for comprehension.
We observed that in-domain participants dedicated more time and effort to explaining their decisions on topics within their field. Notably, 90.9\% of these participants made statements such as `I have learned that in the course' or `this is relevant to our research', indicating a direct connection between their academic training and their approach to the topics. Participants noted several benefits of possessing domain-specific knowledge, including: (i) the ability to quickly fact-check independently, (ii) a deeper understanding of the topic, and (iii) a reduced likelihood of being misled by media or biased narratives.
In comparison, participants without domain knowledge found it challenging to grasp the topics. As a result, they either (i) spent additional time reviewing decision-supporting information or (ii) made decisions without fully understanding the topic and the associated information.


Our findings indicate that in-domain PhD students emphasize domain-specific knowledge, which shapes their focus on the professional aspects of topics rather than the authors' conclusions or opinions. This trend persists even for unfamiliar in-domain topics, where students primarily base their decisions on their interpretation of relevant chemical information in the text.

\subsubsection{How helpful is AIGC for decision-making compared to UGC?}

We asked participants about the role of AIGC in their decision-making process. 
All participants mentioned they detected no traces of AIGC throughout the experiment, suggesting that they were unable to distinguish between AIGC and UGC when provided with incorrect or misleading signals about the information. In particular, participants, including participants in the domain that address topics within their field, assumed the accuracy of the information sources.

Despite recognizing AIGC, 13.63\% of the participants expressed explicit negative attitudes towards its use in decision-making. Participants expressed the reasons for their distrust in AIGC: (i) a general mistrust of generative AI product, (ii) the additional effort required to fact-check AIGC, (iii) a lack of understanding of the generative AI pipeline, which prevents trust, and (iv) previous negative expreiences with generative AI, such as receiving incorrect information or the AI not accurately following prompts. However, there is a noticeable discrepancy between their attitudes and behaviours, as all participants who expressed dislike for AIGC also acknowledged its usefulness in their decision-making. This phenomenon can be explained  (i) they are less suspicious when they see information from a decent non-AI source, and (ii) they did not find any obvious mistakes in AIGC.  Additionally, all three participants believed that their decisions may have differed had they known the actual sources of the information.
We also gathered insights from participants who did not view AIGC negatively for decision-making.  54.55\% of the participants mentioned that AI-generated information as equivalent to human-written information in terms of its utility for decision-making. They attributed their trust in AIGC to several reasons: (i) frequent use of generative AI without  encountering issues, (ii) belief that generative AI vendors like ChatGPT often provide high-quality information that can outperform human output, and (iii) confidence in their ability to make decisions based on the content's informativeness, regardless of its source. Conversely, only one participant explicitly stated that AIGC is never an option for decision-making.

Participants shared their usage patterns for generative AI, revealing varied engagement levels: 4.55\% never use generative AI or lack a strong intention to use it, 36.36\% use it moderately (from bi-weekly to monthly), and 59.10\% use it intensively.
The primary use, reported by 80.95\% of the participants, is for information retrieval,
including (i) accessing news or socially controversial topics and (ii) rough pictures or obtaining preliminary explanations of unfamiliar concepts.
Other uses of AI include
academic purposes (e.g. writing codes, summarising documents, or giving answers to questions), and (ii) daily writing assistance. 
Only 9.52\% of the participants stated that they would ask AI when they wanted information for decision support. It's also noteworthy that even participants who expressed dislike about AIGC in decision-making still utilize AI for other purposes.

\subsubsection{What features make up the most helpful decision-supporting information?}

We aggregated 
findings into two dimensions to explain the participants' core decision-making rationale --- the contextual circumstances and the information quality. The contextual circumstance of the decision refers to participants' personal thoughts and attitudes regarding the topic or matter. 
We found that 86.36\% of the participants were less likely to change their opinions if they had heard of the issue before and had an attitude towards the topic. Participants believe that they found it more challenging to change their thoughts or reject previous decisions if they found the topic familiar or relevant to themselves. They would change their opinions on familiar topics only if (i) the new information is convincing enough (e.g. ground truth from authorised sources), (ii) they admit the previous decision is wrong, or (iii) due to synchronisation pressure from others. 

Furthermore, results suggested that participants designated attitudes towards the topic after a complicated assessment concerning their knowledge, habits, personality traits, customary information consumption, and publicity of that opinion. The contextual factor involves many circumstances, including (i) support the argument that they feel the most reasonable for themselves; (ii) extract the information they can comprehend from the text, then make judgement solely based on common sense, past experience and the extracted parts they can understand; and (iii) support the information from the most authorised and trustable source.

Additionally, 95.45\% of the participants emphasised that information quality is another critical factor affecting decision-making. We tackled information quality from two perspectives by letting participants discuss (i) the features of information that are influential to their decisions and (ii) the features of high-quality information. Information can be influential to decisions, as it can either strengthen or weaken participants' beliefs when the information gives supporting or negative stances. Concerning high information quality, participants mentioned a range of criteria, including (i) obtained from a highly authorised or highly-regarded information source, (ii) stating objective and traceable facts, (iii) there is strong evidence of being fact-checked or validated, and (iv) recommended by trustable personnel from their real lives. Finally, it is worth noting that only one participant admitted that there is interaction between the stances of the information and information quality. For instance, given two pieces of information with the same quality, participants prefer information that supports their existing views. 

\section{Discussion}
The experiment results revealed essential factors that affect participants' online decision-making, including demographic factors like gender, contextual factors like domain-specific knowledge and behavioural factors like thinking time. Among all aspects, gender difference is one of the most addressed factors since last century. Research has been conducted in psychology, sociology, and human-computer interaction (HCI) to investigate whether gender differences affect decision-making \citep{RN48, RN49,RN50,RN76}. Although existing past research tends to believe there exists objective gender differences in decision-making, it is still challenging to assert gender effect on modern online decision-making without a greater-scaled study that fully considers all relevant contextual and demographic features.

\subsection{Domain-Specific Knowledge and Decision-Makings}

From the results of \textbf{RQ1}, domain-specific knowledge played a vital role in decision-making from two perspectives - (i) participants' contextual attitude regarding the topic (existing domain-specific knowledge) and (ii) information retrieval (new domain-specific knowledge intake). Participants' contextual attitude heavily depends on their understanding of the topic. Consider the scenario that multiple participants once faced in this experiment - how to make decisions when there are two pieces of information that both are (i) listing objective facts, (ii) from trustable sources, (iii) extremely challenging to understand without proper education on this field and (iv) showing opposite attitudes and conflicting conclusions. There is no doubt that participants who have domain-specific knowledge would induce more in-depth and valuable thinking. According to the interviews, in-domain PhD students tend to spend more time explicitly explaining their whole reasoning and deduction processes. In contrast, in-domain undergraduate students could conclude their reasons simply as ``I know this because I have once heard it from the lecture". Even though the decision quality is not the primary focus of this research, the differences between participants' reasoning still recall the Dunning-Kruger Effect \citep{RN155}. As \citet{RN155} also implied, there is a high chance that people's acceptance of new professional information increases regardless of whether the information supports or is against the subject's existing ideology. However, it becomes harder to model the subject's opinion change if their domain-specific knowledge level is high as their reasoning complexity also increases. As the domain-specific knowledge level decreases, evidence indicates that information with supporting attitudes is more influential to subjects. 

If subjects without enough domain-specific knowledge encounter in-domain topics, the decision heavily depends on their first impression and the first decision of the topic. It is far-fetched to predict subjects' initial decisions without detailed information about their past activities as the decision could either (i) be made by themselves based on the information they can understand or (ii) depend on the reasoning and conclusions given by the information. Nevertheless, it is feasible to affect their existing opinions. Although subjects tend to believe information from high-quality sources, they still have different information pursuits. Subjects with sufficient domain-specific knowledge can notice the deduction and reasoning logic quality, whereas ordinary subjects are primarily affected by the conclusions and arguments themselves. It leads to a worrying situation because most non-professional subjects' opinion changes predominantly rely on their information retrieval. For those subjects, any information can be helpful for their decision-making as long as it appears scientific, trustworthy, and collected from a seemingly dedicated source. The result indicated that the majority of participants do not constantly have the intention or capability to fact-check the information they retrieved independently, indicating limited self-defence from potentially misleading information regarding topics requiring professional knowledge. Furthermore, the emergence of generative AI-based applications for information retrieval makes the issue more complicated due to its outstanding power to generate `seemingly reliable information' from nowhere automatically \citep{RN299}.

\subsection{Positive and negative roles of AI in modern information retrieval}
Concerning the effect of AIGC on decision-making we observed for \textbf{RQ2}, we can see that generative-AI-based products like ChatGPT have become indispensable for many university students' information retrieval. For a piece of information, being AI-generated is no longer a considerable disadvantage for AIGC during decision-making, even though many participants stated that they are always suspicious about AIGC. The effect of AIGC differs under different circumstances. This work mainly discusses the difference between (i) information labelled as AIGC and (ii) AIGC without labelled sources. Properly sourced AIGC is manageable since all information-related signals are transparent and thus subjects bear their own risk when making decisions. The trouble concerns when AIGC (i) is not sourced appropriately, (ii) pretends to be from a dedicated human-associated source, or (iii) is assumed to be fact-checked by the user. Many participants either do not want to or cannot fact-check the information they receive. Consequently, they lose the opportunity to realise the key background of the information. 

Furthermore, subjects are more vulnerable if they are dealing with topics requiring domain-specific knowledge. As discussed before, subjects tend to draw existing conclusions from the information rather than elaborate on the evidence or reasoning if they have insufficient professional knowledge. The information validation process becomes more costly as the topics become more domain-specific. Hence, it brings up new challenges to gatekeep users from the potential negative influence of AIGC. However, on the positive side, existing research discussed the potential of AIGC to provide easier-to-understand explanations to users. Besides, generative-AI-based information retrieval shows irreplaceable convenience and feasibility. Although this work only discussed a particular aspect of the potential bias of AIGC for decision-making, we are still confident that AI has vast potential for unbiased information retrieval once appropriate regulation and moderation method is investigated and implemented.

\lsq{\subsection{AI-involved online decision-making frameworks}}

\lsq{The results demonstrate that AIGC can be as useful as human-written information, even when both are cited from appropriate sources, regardless of the user's domain-specific literacy. Therefore, AI-assisted online decision-making can be examined from two complementary perspectives: technology adoption and cognitive processing.}

\lsq{From the technology-adoption perspective, it is essential to understand how users perceive AI and how they incorporate it into information retrieval and subsequent decision-making. The experimental results indicate that users are more likely to rely on AI when they do not hold a strong aversion to it, which aligns with the Technology Acceptance Model (TAM) proposed by \citet{RN338, RN339}. TAM suggests that technology adoption depends primarily on perceived usefulness (the degree to which the user believes the technology enhances task performance) and perceived ease of use (the degree of effort required to learn and operate the system). Despite the relatively low perceived ease of use commonly associated with generative-AI applications, our findings suggest that perceived usefulness becomes the dominant factor in decision-making scenarios that require domain-specific knowledge. However, perceived usefulness is not static; it varies across individuals and task domains, and may shift over time alongside the rapid evolution of large language models. It can also fluctuate with changes in the user's domain-specific literacy. As suggested by the Dunning–Kruger effect \citep{RN340}, individuals' confidence and reasoning strategies shift as their knowledge develops. Therefore, to accurately model AI’s influence on decision-making, it is necessary to establish a systematic and objective approach for measuring perceived usefulness over time.}

\lsq{From the cognitive-processing perspective, AI-assisted decision-making can be analysed using the Elaboration Likelihood Model (ELM). ELM proposes two cognitive pathways in information processing: a central route, involving effortful reasoning and critical evaluation, and a peripheral route, based on heuristics and superficial cues. Decisions formed via the central route tend to be more stable and resistant to change, whereas those formed via the peripheral route are made more easily but are more susceptible to revision. This framework aligns with both our quantitative and qualitative results: individuals are more likely to engage in central-route processing when (i) they perceive the topic as personally relevant, (ii) they possess sufficient familiarity to discuss it in depth, and (iii) they can draw upon domain-specific knowledge to demonstrate their expertise. As noted by \citet{RN341}, individuals’ information preferences in decision-making depend on which cognitive route is activated. Thus, if users’ cognitive pathways can be predicted in advance, it becomes feasible to design more effective strategies for using AI to support or persuade users. In this sense, future large language models and generative-AI systems hold substantial potential for persuasive and advisory tasks, such as consulting, education, and healthcare.}

\subsection{Limitations and future research}

Our study is not without limitations, and we considered both internal and external validity \cite{wohlin2012experimentation}. \lsq{The first limitation is about sample demographics. This research solely focuses on university students' human-computer interaction and decision-making behaviours. However, the recruitment may introduce potential sample bias because we failed to consider a wider range of sample demographics. Therefore, recommend future studies to expand the expected sample pool to consider a wider range of cultural, social and language factors to improve generalisability. }

\lsq{The second limitation is about the topic domains covered in this research. This study deployed a single comparison group of non-chemical background versus chemical background to examine the effect of domain-specific knowledge. However, it is recommended for future studies to expand the scope to look into a wider spectrum of domains such as law, social science, semiconductors and more. In this way, we can obtain additional practical and empirical evidence. }

\lsq{Thirdly, we failed to attach a systematic quality control and evaluation pipeline for AIGC in this research. This may cause potential bias if AIGC is not under the researcher's complete control. Thus, for future studies, we believe it is necessary to deploy a fully objective quality assessment process of AIGC to avoid potential ambiguity and strengthen the dominion of AIGC.}

 Despite these limitations, we believe that our methodological approach, combining surveys with think-aloud interviews, offers valuable insights into how domain-specific knowledge and the AIGC influcen participant decision-making process. This foundation sets the stage for future studies to further explore these dynamics.

\section{Conclusion}

In this research, we conducted a lab-based explanatory study to investigate (i) factors associated with decision-making and opinion change, (ii) the effect of domain-specific knowledge and (iii) the influence of information source (e.g. whether or not being AIGC). The result indicated that participants' decision-making is profoundly affected by their familiarity with the topic and the domain-specific knowledge. Participants are less likely to change opinions on topics that require domain-specific knowledge due to the lack of capability to understand the arguments and evidence. Furthermore, we found that AIGC has become essential for people's daily information retrieval and decision-making. There is huge potential for AI to become a key instrument for people to overcome the barrier of domain-specific knowledge once the bias problem and other potential issues are properly addressed and regulated.
\section{CRediT authorship contribution statement}

\textbf{Shangqian Li:} Writing - Original Draft, Writing - Review \& Editing, Visualization, Conceptualization, Methodology, Software, Validation, Formal analysis, Investigation, Resources, Data Curation. \textbf{Tianwa Chen:} Writing - Original Draft, Writing - Review \& Editing, Supervision. \textbf{Gianluca Demartini:} Writing - Review \& Editing, Supervision, Project administration, Funding acquisition.

\section{Data available statement}
The data that support the findings of this study are available from the corresponding author upon reasonable request.

\section{Conflict of interest statement}
None of the authors have a conflict of interest to disclose

\section{Funding statement}

Research supported by:

Schweizerischer Nationalfonds zur Förderung der Wissenschaftlichen Forschung (10.13039/501100001711)

\* CRSII5\_205975













\newpage
\bibliographystyle{elsarticle-harv} 
\bibliography{bib}





\end{document}